\documentclass[twocolumn,aps,epsf,graphics,psfig]{revtex4}
\usepackage{graphicx,graphics} % Include figure files
\usepackage[dvips,bookmarks=false]{hyperref}

\usepackage{graphicx}
\usepackage{dcolumn}

\usepackage{eucal}
\usepackage[dvips]{epsfig}
\begin{document}
\title{Tunnel Magnetoresistance of a Single-Molecule Junction}
\author{Alireza Saffarzadeh$^{1,2,}$}
\altaffiliation{E-mail: a-saffar@tehran.pnu.ac.ir}
\affiliation{$^1$Department of Physics,
Payame Noor University, Nejatollahi Street, 159995-7613 Tehran, Iran \\
$^2$Computational Physical Sciences Laboratory, Department of
Nano-Science, Institute for Research in Fundamental Sciences
(IPM), P.O. Box 19395-5531, Tehran, Iran}
\date{\today}

\begin{abstract}
Based on the non-equilibrium Green's function (NEGF) technique and
the Landauer-B\"{u}ttiker theory, the possibility of a molecular
spin-electronic device, which consists of a single C$_{60}$
molecule attached to two ferromagnetic electrodes with finite
cross sections, is investigated. By studying the coherent
spin-dependent transport through the energy levels of the
molecule, it is shown that the tunnel magnetoresistance (TMR) of
the molecular junction depends on the applied voltages and the
number of contact points between the device electrodes and the
molecule. The TMR values more than 60\% are obtained by adjusting
the related parameters.

\end{abstract}
\maketitle

%{\bf PACS.}
\section{Introduction}
Traditional magnetic tunnel junctions use inorganic insulators as
spacers \cite{Moodera1,Miyazaki}. However, recent experimental
data have clearly shown that organic molecules can serve the same
purpose and a rather large TMR can be found
\cite{Tsukagoshi,Xiong,Dediu,Shim,Santos,Ouyang,Petta}. In these
experiments, spin polarized transport through molecular layers
sandwiched between two magnetic layers has been demonstrated for
systems involving carbon nanotubes \cite{Tsukagoshi}, other
organic $\pi$ conjugate systems \cite{Xiong,Dediu,Shim,Santos},
molecular bridges \cite{Ouyang}, and self-assembled organic
monolayers \cite{Petta}. Organic materials have relatively weak
spin-orbit interaction and weak hyperfine interaction, so that
spin memory can be as long as a few seconds \cite{Sanvito}. Such
features make them ideal for spin-polarized electron injection and
transport applications in molecular spintronics.

Spin-polarized transport through organic molecules sandwiched
between two magnetic contacts has also been recently investigated
theoretically
\cite{Pati,Senapati,Dalgleish,Liu,Rocha,Waldron,Wang,He,Ning,Wang2,Mehrez,Krompiewski,Emberly}.
Many of these calculations, based on density functional theory or
tight-binding model, have shown that by changing the magnetic
alignment of the contacts one can substantially affect the
electronic current in the molecular devices. For instance, in a
single benzene-1-4-dithiolate molecule sandwiched between two Ni
clusters, parallel magnetic alignment led to significantly higher
current (about one order of magnitude) than antiparallel
alignment, suggesting the possibility of a molecular spin valve
\cite{Pati}.

Among many types of molecules, the fullerene C$_{60}$ which is one
of the most well-investigated organic semiconductors, is suitable
as a molecular bridge in magnetic tunnel junctions, because its
lowest unoccupied molecular orbital (LUMO) is situated at
relatively lower energies in comparison with the other organic
molecules. In recent years, the electrical and magnetic properties
of C$_{60}$-Co nanocomposites, where Co nanoparticles are
dispersed in C$_{60}$ molecules have been studied
\cite{Zare,Sakai,Miwa}, and the maximum TMR ratio of about 30\% at
low bias voltages was reported. Recently, He \textit{et al.}
\cite{He2}, studied the spin-polarized transport in Ni/C$_{60}$/Ni
junction using density functional theory and the
Landauer-B\"{u}ttiker formalism. They showed that the binding
sites of Ni on C$_{60}$ molecule play a crucial role in the
transport properties of the system and a large value for junction
magnetoresistance was predicted. We believe, however, that no
theoretical study on spin-polarized transport through a single
C$_{60}$ molecule and its TMR effect, based on tight-binding
method, has so far been reported.
\begin{figure}
\centerline{\includegraphics[width=0.9\linewidth]{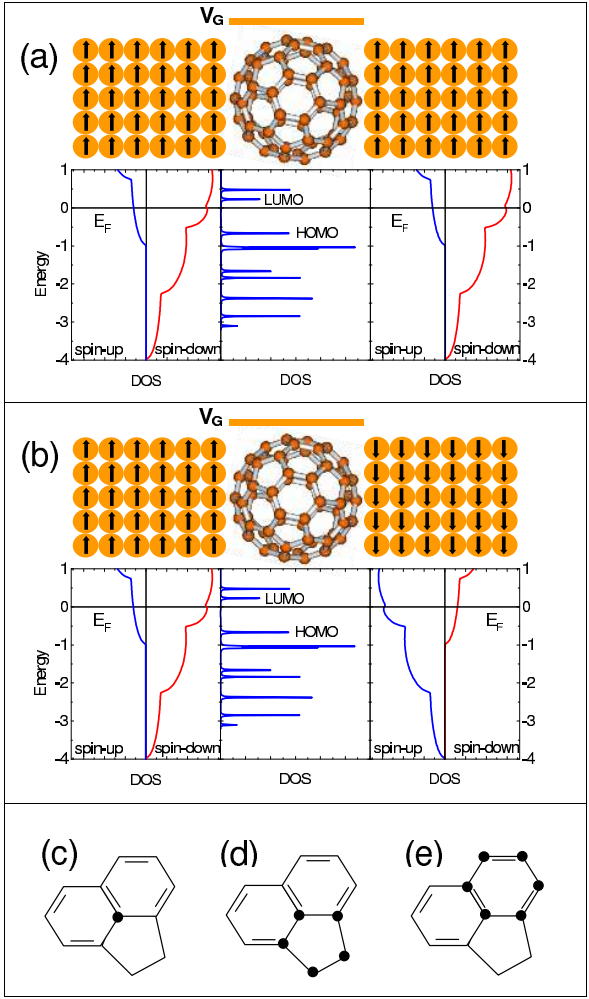}}
\caption{Schematic view of the FM/C$_{60}$/FM molecular junction
at zero voltages for (a) parallel, and (b) antiparallel alignments
of the magnetizations in the FM electrodes. (c)-(e) The three
different ways of coupling between the C$_{60}$ molecule and the
magnetic electrodes used in this work. Black circles show the
position and the number of couplings.}
\end{figure}

In this paper, we use a single C$_{60}$ molecule in between two
ferromagnetic (FM) electrodes, instead of the usual insulator
layer as in the standard TMR setup, to investigate the possibility
of a single-molecule spintronics device. An interesting feature of
the C$_{60}$ molecule is the emergence of a quantum loop current
which is related to the degeneracy of the energy levels of the
molecule and its magnitude can be much larger than that of the
source-drain current \cite{Naka}. This feature is useful for
spin-polarized transport through the molecule. Our approach
indicates that the spin currents in such a molecular junction are
mainly controlled by the molecular field of the FM electrodes, the
electronic structure of the molecule and its coupling to the
electrodes.

\section{Model and formalism}
We calculate the spin currents through a single C$_{60}$ molecule,
sandwiched between two semi-infinite FM electrodes with simple
cubic structure and square cross section ($x$-$y$ plane). The
model of such a structure is shown schematically in Fig. 1(a) and
1(b). Since the electron conduction is mainly determined by the
central part of the junction, the electronic structure of this
part should be resolved in detail. It is therefore reasonable to
decompose the total Hamiltonian of the system as
\begin{equation}
\hat{H}=\hat{H}_{L}+\hat{H}_{C}+\hat{H}_{R}+\hat{H}_{T}\  .
\end{equation}

The Hamiltonian of the left ($L$) and right ($R$) electrodes is
described within the single-band tight-binding approximation and
is written as
\begin{equation}
\hat{H}_{\alpha}=\sum_{i_\alpha,\sigma}(\epsilon_{0}-\mathbf{\sigma}\cdot
\mathbf{h}_{\alpha})\hat{c}_{i_\alpha,\sigma}^\dag\hat{c}_{i_\alpha,\sigma}-\sum_{<i_\alpha,
j_\alpha>,\sigma}t_{i_\alpha,j_\alpha}\hat{c}_{i_\alpha,\sigma}^\dag\hat{c}_{j_\alpha,\sigma}\
,
\end{equation}
where $\hat{c}_{i_\alpha,\sigma}^\dag$
($\hat{c}_{i_\alpha,\sigma}$) creates (destroys) an electron with
spin $\sigma$ at site $i$ in electrode $\alpha$ (=$L$, $R$), the
hoping parameter $t_{i_\alpha,j_\alpha}$ is equal to $t$ for the
nearest neighbors and zero otherwise. Here, $\epsilon_{0}$ is the
spin independent on-site energy and will be set to $3t$ as a shift
in energy, $-\mathbf{\sigma}\cdot \mathbf{h}_{\alpha}$ is the
internal exchange energy with $\mathbf{h}_{\alpha}$ denoting the
molecular field at site $i_\alpha$, and $\mathbf{\sigma}$ being
the conventional Pauli spin operator. The Hamiltonian of the
C$_{60}$ molecule in the absence of FM electrodes is expressed as
\begin{equation}
\hat{H}_{C}=\sum_{i_C,\sigma}\epsilon_{i_C}\hat{d}_{i_C,\sigma}^\dag\hat{d}_{i_C,\sigma}-
\sum_{<i_C,j_C>,\sigma}t_{i_C,j_C}\hat{d}_{i_C,\sigma}^\dag\hat{d}_{j_C,\sigma}\
,
\end{equation}
where $\hat{d}_{i_C,\sigma}^\dag$ ($\hat{d}_{i_C,\sigma}$) creates
(destroys) an electron with spin $\sigma$ at site $i$ of C$_{60}$
and $\epsilon_{i_C}$ is the on-site energy and will be set to zero
except in the presence of gate voltage $V_G$ that shifts the
energy levels of the molecule and hence $\epsilon_{i_C}=V_G$. The
hoping strength $t_{i_C,j_C}$ in C$_{60}$ molecule depends on the
C-C bond length; thus, we assume different hoping matrix elements:
$t_1$ for the single bonds and $t_2$ for the double bonds. Most
recently, we have shown that the effect of bond dimerization may
considerably affect the electron conduction through the molecule
under suitable conditions \cite{Saffar1}. Finally, $\hat{H}_T$
describes the coupling between the FM leads and the molecule and
takes the form
\begin{equation} \hat{H}_{T}=-\sum_{\alpha=\{L,R\}}\sum_{i_\alpha,
j_C,\sigma}t_{i_\alpha,j_C}(\hat{c}_{i_\alpha,\sigma}^\dag\hat{d}_{j_C,\sigma}+\textrm{H.c.})\
.
\end{equation}
The hopping elements $t_{i_\alpha,j_C}$ between the lead orbitals
and the $\pi$ orbitals of the molecule are taken to be $t'$. In
this study we assume that the electrons freely propagate and the
only resistance arising from the contacts. This means that the
transport is ballistic \cite{Datta}; therefore, we set $t'=0.5 t$,
because the value of $t'$ should not be smaller than the order of
$t$. On the other hand, we assume that the spin direction of the
electron is conserved in the tunneling process through the
molecule. Therefore, there is no spin-flip scattering and the
spin-dependent transport can be decoupled into two spin currents:
one for spin-up and the other for spin-down. This assumption is
well-justified since the spin diffusion length in organics is
about 4 nm \cite{Sanvito} and especially in carbon nanotubes is at
least 130 nm \cite{Tsukagoshi}, which are greater than the
diameter of C$_{60}$ molecule ($\sim$ 0.7 nm).

Since the total Hamiltonian does not contain inelastic
scatterings, the spin currents for a constant bias voltage, $V_a$,
are calculated by the Landauer-B\"{u}ttiker formula based on the
NEGF method \cite{Datta}:
\begin{equation}\label{I}
I_\sigma(V_a)=\frac{e}{h}\int_{-\infty}^{\infty}
T_\sigma(\epsilon,V_a)[f(\epsilon-\mu_L)-f(\epsilon-\mu_R)]d\epsilon
\ ,
\end{equation}
where $f$ is the Fermi distribution function, $\mu_{L,R}=E_F\pm
\frac{1}{2}eV_a$ are the chemical potentials of the electrodes,
and $T_\sigma(\epsilon,V_a)=\mathrm{Tr}[\hat{\Gamma}_{L,\sigma}
\hat{G}_\sigma \hat{\Gamma}_{R,\sigma}\hat{G}_\sigma^{\dagger}]$
is the spin-, energy- and voltage-dependent transmission function.
The spin-dependent Green's function of the C$_{60}$ molecule
coupled to the two FM electrodes (source and drain) in the
presence of the bias voltage is given as
\begin{equation}
\hat{G}_\sigma(\epsilon,V_a)=[\epsilon \hat{1}
-\hat{H}_C-\hat{\Sigma}_{L,\sigma}(\epsilon-eV_a/2)
-\hat{\Sigma}_{R,\sigma}(\epsilon+eV_a/2)]^{-1}\  ,
\end{equation}
where $\hat{\Sigma}_{L,\sigma}$ and $\hat{\Sigma}_{R,\sigma}$
describe the self-energy matrices which contain the information of
the electronic structure of the FM electrodes and their coupling
to the molecule. These can be expressed as $
\hat{\Sigma}_{\alpha,\sigma}(\epsilon)=
\hat{\tau}_{C,\alpha}\hat{g}_{\alpha,\sigma}(\epsilon)\hat{\tau}_{\alpha,C}$
where $\hat{\tau}$ is the hopping matrix that couples the molecule
to the leads and is determined by the geometry of the
molecule-lead bond. $\hat{g}_{\alpha,\sigma}$ are the surface
Green's functions of the uncoupled leads i.e., the left and right
semi-infinite magnetic electrodes, and their matrix elements are
given by
\begin{equation}\label{g}
g_{\alpha,\sigma}(i,j;z)=\sum_{\mathbf{k}}\frac{\psi_\mathbf{k}(\mathbf{r}_i)\psi^*_\mathbf{k}(\mathbf{r}_j)}
{z-\epsilon_0+\mathbf{\sigma}\cdot
\mathbf{h}_{\alpha}+\varepsilon(\mathbf{k})}\  ,
\end{equation}
where $\mathbf{r}_i\equiv(x_i,y_i,z_i)$,
$\mathbf{k}\equiv(l_x,l_y,k_z)$, $z=\epsilon+i\delta$,
\begin{equation}
\psi_\mathbf{k}(\mathbf{r}_i)=\frac{2\sqrt{2}}{\sqrt{(N_x+1)(N_y+1)N_z}}
\sin(\frac{l_xx_i\pi}{N_x+1})\sin(\frac{l_yy_i\pi}{N_y+1})\sin(k_zz_i)\
,
\end{equation}
and
\begin{equation}
\varepsilon(\mathbf{k})=2t[\cos(\frac{l_x\pi}{N_x+1})+
\cos(\frac{l_y\pi}{N_y+1})+\cos(k_za)]\ .
\end{equation}

Here, $l_{x,y}$ ($=1,...,N_{x,y}$) are integers,
$k_z\in[-\frac{\pi}{a},\frac{\pi}{a}]$, and $N_\beta$ with $\beta
=x, y, z$ is the number of lattice sites in the $\beta$ direction.
Note that $N_x$ and $N_y$ correspond to the number of atoms at the
cross-section of the FM electrodes. Using
$\hat{\Sigma}_{\alpha,\sigma}$, the coupling matrices
$\hat{\Gamma}_{\alpha,\sigma}$, also known as the broadening
functions, can be expressed as
$\hat{\Gamma}_{\alpha,\sigma}=-2\mathrm{Im}(\hat{\Sigma}_{\alpha,\sigma})$.

\begin{figure}
\centerline{\includegraphics[width=0.9\linewidth]{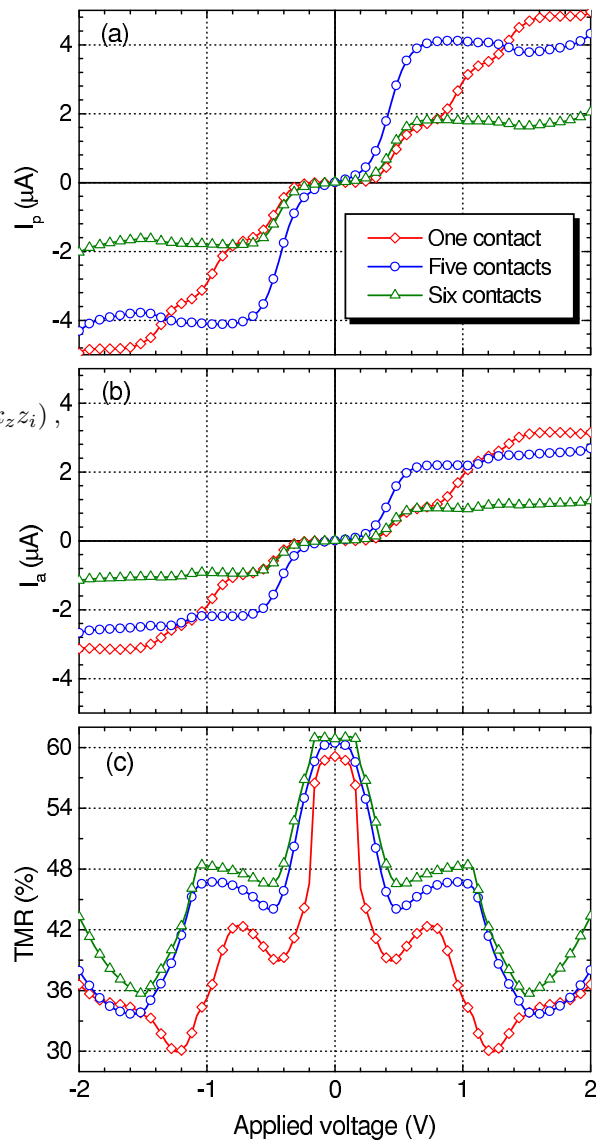}}
\caption{The total currents at $V_G$=0.0 V for the parallel (a)
and antiparallel (b) alignments of magnetizations in the FM
electrodes, and (c) the TMR as a function of applied voltage.}
\end{figure}

In the semi-infinite FM electrodes described by the single-band
tight-binding model, only the central site at the cross section is
connected to the molecule. Furthermore, when the molecule is
brought close to an electrode, the bonding between them will
depend on the molecule orientation. This orientation can be such
that only one carbon atom, a pentagon or a hexagon of the C$_{60}$
molecule be in contact with the leads [see Figs. 1(c)-1(e)].
Therefore, one can expect different conduction through the
molecule, which arises due to the resonant tunneling and the
quantum interference effects. Our approach, as a real-space
method, makes it possible to model arbitrarily the number of
contacts. In this regard, the core of the problem lies in the
calculation of the spin-dependent self-energies
$\hat{\Sigma}_{L,\sigma}$ and $\hat{\Sigma}_{R,\sigma}$. In the
case of contact through a single carbon atom of the molecule, only
one element of the self-energy matrices is non-zero. However, for
the transport through opposite pentagons or hexagons, 25 or 36
elements of the self-energy matrices are non-zero, respectively.

The total charge current is given by $I=I_\uparrow+I_\downarrow$.
In this case, we calculate the TMR ratio from the usual
definition: $\mathrm{TMR}\equiv(I_p-I_a)/I_p$, where $I_{p,a}$ are
the total currents in the parallel and antiparallel alignments of
magnetizations in the FM electrodes, respectively.

\section{Results and discussion}
We now use the method described above to study the coherent
spin-dependent transport and magnetoresistance effect of
FM/C$_{60}$/FM molecular junction. We have done the numerical
calculations for the case that the direction of magnetization in
the left FM electrode is fixed in the +$y$ direction, while the
magnetization in the right electrode is free to be flipped into
either the +$y$ or -$y$ direction. We set
$|\mathbf{h}_{\alpha}|$=1.5 eV, $t$=1 eV, $t_1=t$, $t_2=1.1\,t$,
$N_x=N_y=5$, and $T$=300 K in the calculations.

In Fig. 2 we show the current-voltage ($I$-$V$) characteristics
(in the parallel and antiparallel alignments) and also the TMR as
a function of applied voltage for three different ways of coupling
between the C$_{60}$ molecule and the FM electrodes. These three
cases were chosen as the most probable experimental orientations
\cite{Hashizume}. As can be seen, the $I$-$V$ curves show a
steplike behavior which indicates that a new channel is opened.
The values of currents in the parallel configuration are nearly
two times larger than that in the case of antiparallel one. The
difference between $I_p$ and $I_a$ is related to the asymmetry of
surface density of states (SDOS) of the FM electrodes for spin-up
and spin-down electrons [see Figs. 1(a) and 1(b)]\cite{Explain},
and the quantum tunneling phenomenon through the molecule. Here,
we have assumed that only itinerant $\textit{sp}$-like electrons
contribute to the tunneling current, which is reasonable when the
distance between the two FM electrodes is greater than 0.5 nm
\cite{Munzenberg}.

In the parallel alignment, minority electrons go into the minority
states by tunneling through the molecule and there is no asymmetry
in the SDOS. If, however, the two FM electrodes are magnetized in
opposite directions, the minority (majority) electrons from the
left electrode seek empty majority (minority) states in the right
electrode, that is, asymmetry in the SDOS. In fact, in the
antiparallel alignment, the tunneling currents for both spin
channels are asymmetric with respect to voltage inversion (not
shown). Consequently, the parallel arrangement gives much higher
total current through the C$_{60}$ molecule than does the
antiparallel arrangement. This difference in the total currents is
the origin of TMR effect which has been shown in Fig. 2(c). The
TMR ratio has it maximum value (more than 60\%) at low bias
voltages. With increasing the applied voltage, we first observe
that the TMR decreases. Such a behavior is similar to the
conventional magnetic tunnel junctions \cite{Moodera3}. With
further increase in the bias voltage, the TMR ratio increases and
then, the reduction and enhancement of TMR are repeated. Recently,
using C$_{60}$-Co nanocomposites, Miwa {\it et al}. \cite{Miwa}
measured a magnetoresistance value as large as 18\% due to the
spin-polarized tunneling of carriers between Co nanoparticles via
C$_{60}$ molecules. Our bias-voltage dependence of TMR is
qualitatively in agreement with their results. A similar
bias-dependent behavior in which TMR ratio increases with applied
voltage, using a EuS spin-filter tunnel barrier in conventional
junctions, has also been reported \cite{Nagahama2}.
\begin{figure}
\centerline{\includegraphics[width=0.9\linewidth]{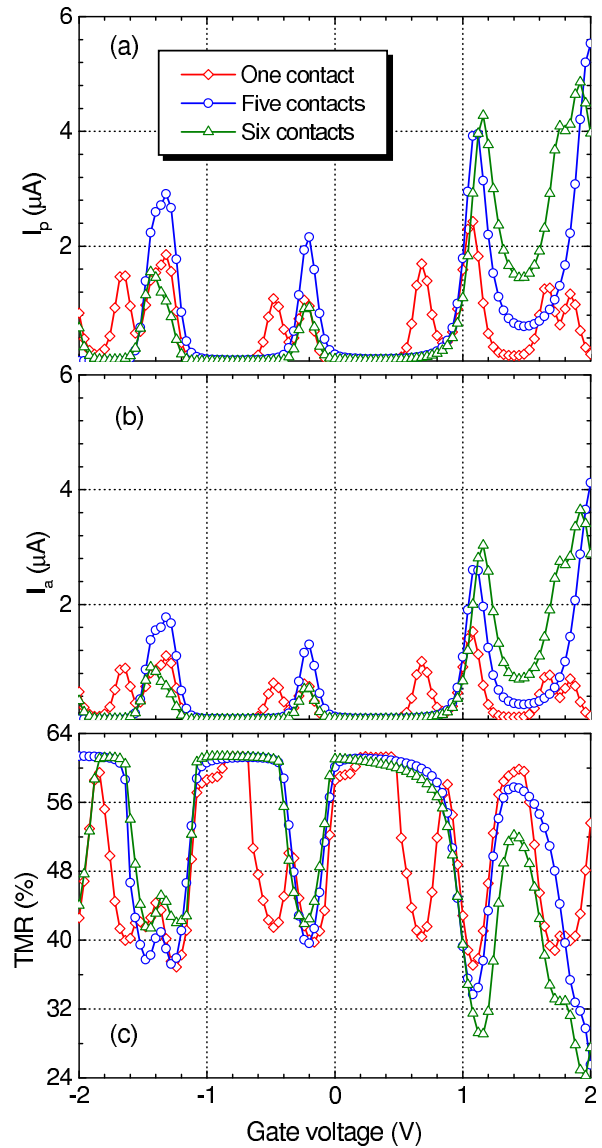}}
\caption{The total currents at $V_a$=0.1 V for the parallel (a)
and antiparallel (b) alignments of magnetizations in the FM
electrodes, and (c) the TMR as a function of gate voltage.}
\end{figure}

In order to investigate the other features of the junction, we
show in Fig. 3 the effects of gate voltage on the total currents
and the TMR in the cases of single and multiple contacts. When the
gate voltage is zero, we showed in Fig. 2(c) that the maximum
values for the TMR are obtained at low applied voltages ($V_a\leq
0.1$ V). In such cases, the number of states contained in the
energy window between $\mu_L$ and $\mu_R$ is zero, because the
energy window lies in the highest occupied molecular orbital
(HOMO)-LUMO gap of the C$_{60}$ molecule [Figs. 1(a) and 1(b)],
where there is no molecular level. Therefore, the current flow
mechanism is tunneling. For this reason, the voltage dependence of
the TMR effect at low voltages is similar to that of the
conventional magnetic tunnel junctions. It is worth mentioning
that in the selected voltage interval, we did not observe
significant changes in the transmission spectra with increasing
the bias voltage.

Applying a gate voltage shifts the molecular levels relative to
the Fermi level of electrodes, and hence the transmission
coefficients may significantly vary \cite{Saffar1}. Since the
currents depend on the molecular density of states lying between
$\mu_L$ and $\mu_R$, then, by increasing the gate voltage, HOMO or
LUMO peak moves inside the energy window and the total currents
$I_p$ and $I_a$ increase. In this case, for either negative or
positive gate voltages, the current flow mechanism is resonant
tunneling, and a peak appears in the current curves. Figures 3(a)
and 3(b) show that the currents vanish in a wide range of gate
voltages, because in these voltages there is no resonant level
inside the energy window. In both figures, the peaks appear at the
same voltages which confirm that the appearance of current peaks
is due to the molecular levels. However, the height of peaks is
different which is due to the difference in the spin-dependent
SDOS of the FM electrodes. The results suggest that the C$_{60}$
molecule is an interesting candidate for operation of devices as a
nano-scale current switch. Also, with increasing the number of
contact points between the device electrodes and the molecule, the
interference effects around these points become important, some
resonances might completely disappear, and the spin current
changes. This is the reason of difference between the currents in
the single and multiple contacts [see Figs. 2 and 3]
\cite{Saffar1,Paul,Palacios}.

\section{Conclusion}
Using the NEGF method and the Landauer-B\"{u}ttiker theory, we
have investigated the possibility of making a C$_{60}$-based
magnetic tunnel junction. We have shown that coupling between the
single C$_{60}$ molecule and magnetic electrodes in FM/C$_{60}$/FM
structure, produces large magnetoresistance effects (greater than
60\%) that can be modulated by using the molecule orientation,
bias and gate voltages to control the spin-polarized transport.
The present study advances the fundamental understanding of
spin-dependent transport in molecular junctions and suggests that
the C$_{60}$ molecule is an interesting candidate for application
in the magnetic memory cells and spintronic devices.

Throughout this study, we have ignored the effects of inelastic
scattering and the magnetic anisotropy of the FM electrodes due to
the reduced dimensionality and the geometry of the system. These
factors can affect the spin-dependent transport and hence, another
improved approach is needed for more accurate results.

\section*{Acknowledgement}
The author thanks Professor J.S. Moodera for valuable comments.
This work was supported by Payame Noor University grant.

\end{document}